\documentclass[a4paper]{article}

\usepackage{icrc2013}

\title{The Real-Time Analysis of the Cherenkov Telescope Array Observatory}

\shorttitle{The Real-Time Analysis of the CTA Observatory}

\authors{
A. Bulgarelli$^{1}$, V. Fioretti$^{1}$, J.L. Contreras$^{3}$, A. Lorca$^{3}$, A. Aboudan$^{17}$, J. J. Rodr\'iguez-V\'azquez$^{16}$,  S. Lombardi$^{7}$, G.Maier$^{4}$, L. A. Antonelli$^{7}$, D. Bastieri$^{8}$, C. Boisson$^{9}$, J. Borkowski$^{10}$, S. Buson$^{8}$, A. Carosi$^{7}$, V. Conforti$^{1}$, A. Djannati-Ata\"{\i}$^{11}$, J. Dumm$^{12}$, P. Evans$^{5}$,  L. Fortson$^{12}$, F. Gianotti$^{1}$, R. Graciani$^{13}$, P. Grandi$^{1}$, J. Hinton$^{5}$, B. Humensky$^{14}$, K. Kosack$^{15}$, G. Lamanna$^{2}$,   G. Malaguti$^{1}$, M. Marisaldi$^{1}$, L. Nicastro$^{1}$, S. Ohm$^{5}$, J. Osborne$^{5}$,  S. Rosen$^{5}$, M. Trifoglio$^{1}$, G. Tosti$^{6}$
for the CTA Consortium.
}

\afiliations{
$^1$ INAF/IASF Bologna, Bologna, Italy \\
$^2$ Laboratoire d'Annecy-le-Vieux de Physique des Particules, Universit\'e de Savoie, CNRS/IN2P3, France \\
$^3$ Dept. FAMN, Universidad Complutense de Madrid, Grupo de Altas Energ\'{\i}as, Spain \\
$^4$ Deutsches Elektronen-Synchrotron, Platanenallee 6, Zeuthen, Germany \\
$^5$ Dept. of Physics and Astronomy, University of Leicester, United Kingdom \\
$^6$ Universit\`a di Perugia, Italy \\
$^7$ INAF - Osservatorio Astronomico di Roma, Italy \\
$^8$ Dipartimento di Fisica, Universit\'a degli Studi di Padova, Italy \\
$^9$ Observatoire de Paris, LUTH, CNRS, Universit\'e Paris Diderot, France \\
$^{10}$ Copernicus Astronomical Center, Polish Academy of Sciences, Poland \\
$^{11}$ AstroParticule et Cosmologie, France \\
$^{12}$ University of Iowa, Department of Physics and Astronomy, USA \\
$^{13}$ Departament d'Astronomia i Meteorologia, Institut de Ci\`encies del Cosmos, Universitat de Barcelona, Spain \\
$^{14}$ Department of Physics and Astronomy, Barnard College; Department of Physics, Columbia University, USA \\
$^{15}$ CEA/DSM/IRFU, Saclay, France \\
$^{16}$ Centro de Investigaciones Energ\'eticas, Medioambientales y Tecnol\'ogicas,  Madrid, Spain\\
$^{17}$ CISAS, University of Padua,  Padua, Italy

}

\email{bulgarelli@iasfbo.inaf.it}

\abstract{The Cherenkov Telescope Array (CTA) Observatory must be capable of issuing fast alerts on variable and transient sources to maximize the scientific return. This will be accomplished by means of a Real-Time Analysis (RTA) pipeline, a key system of the CTA observatory. The latency and sensitivity requirements of the alarm system impose a challenge because of the large foreseen data flow rate, between 0.5 and 8 GB/s. As a consequence, substantial efforts toward the optimization of this high-throughput computing service are envisaged, with the additional constraint that the RTA should be performed on-site (as part of the auxiliary infrastructure of the telescopes). In this work, the functional design of the RTA pipeline is presented. }


\def \gray {$\gamma$-ray }
\def \aa {A$\&$A$\;$}

\begin{document}
\maketitle

\section{Introduction}

About one-third of the sources contained in the current catalogues of EGRET \cite{bib:Hartman_1999}, AGILE \cite{bib:Pittori_2009}  and Fermi \cite{bib:Nolan_2012} have no known or confirmed counterparts at other wavelengths. The variability aspect is a key factor for understanding these unknown \gray sources; also any active state of a well-known \gray source can trigger the generation of a scientific alert. In addition, the understanding of the radiation mechanisms of the \gray emission during active states, e.g. from High Mass X-ray Binaries, remains puzzling at the best. 

To maximize the science return on time-variable and transient phenomena the Cherenkov Telescope Array (CTA) Observatory  \cite{bib:CTA} must be capable of issuing alerts of unexpected events from astrophysical sources to other facilities. To achieve this goal a Real-Time Analysis (RTA) system is currently being designed.

\begin{figure}[t]
  \centering
  \includegraphics[width=0.48\textwidth]{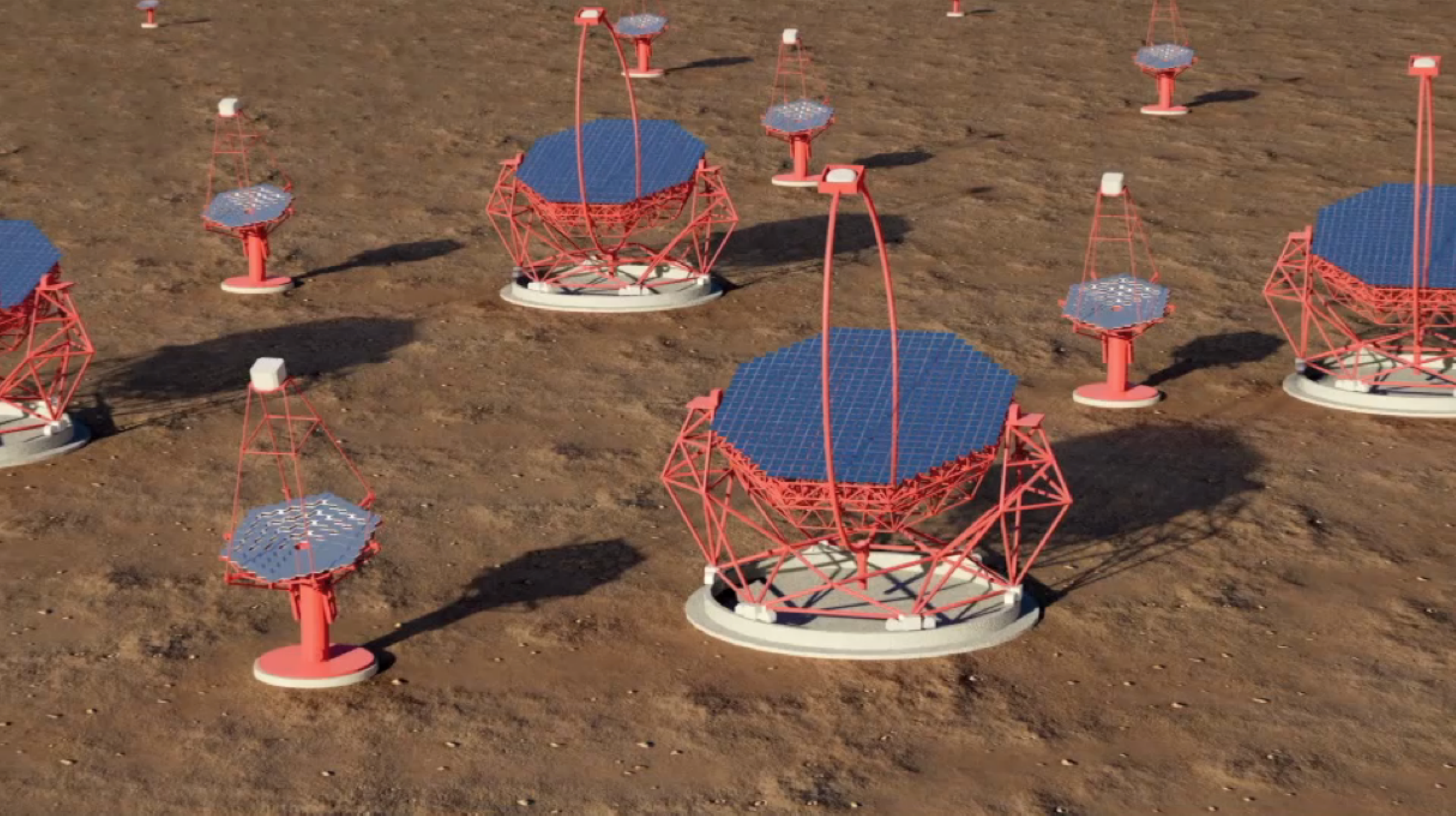}
  \caption{An artist impression of the Cherenkov Telescope Array. Source G. Perez, Instituto de Astrofísica de Canarias, Servicio Multimedia (SMM/IAC).}
  \label{CTA_fig}
 \end{figure}

CTA will be the biggest ground-based very-high-energy (VHE) \gray observatory. Among many outstanding features, CTA will achieve a factor of 10 improvement in sensitivity from some tens of GeV to beyond 100 TeV with respect to existing telescopes. In order to scan the whole sky, CTA will consist of two arrays: one in the south hemisphere, to observe the wealth of sources in the central region of our Galaxy, and one in the north, primarily devoted to the study of Active Galactic Nuclei (AGN) \cite{bib:zech}, galaxies at cosmological distances, and star formation and evolution. To accomplish the science goals, tens of telescopes (Figure \ref{CTA_fig}) of three different sizes will be required: a Large Size Telescope for the lowest energies (20 GeV - 1 TeV), a Medium Size Telescope for the 100 GeV - 10 TeV energy domain, and the Small Size Telescope for the highest energies (few TeV -  beyond 100 TeV).

Listed among the priority projects on the roadmaps of several European networks (ESFRI, ASPERA/ApPEC, ASTRONET) and by the US 2010 Decadal Survey, CTA has received funding from the FP7 program. Currently in the preparatory phase (2011-2014), at the time of writing the international CTA consortium counts more than 1000 scientists from 27 countries.  

In Section~2  we present the general overview of the RTA and its requirements, technological constraints and the observatory operations.  In Section~3 we describe why this is a key system for the CTA observatory. In Section~4 we describe a functional view of the system. Section~5 contains an overview of a technological assessment useful to understand which software and hardware architecture are able to cope with the sustainability of the foreseen CTA data rate. 

\section{A complex Science Alert System}

With its large detection area, CTA will resolve flaring and time-variable emission on sub-minute time scales. Current instruments have already detected flares varying on time scales of a few minutes \cite{bib:min1, bib:min2}. CTA will be 10000 times more sensitive to flares than Fermi at 25 GeV, and combining both sites will achieve full sky coverage. Current CTA simulations show that extreme AGN outbursts, which in the past have reached flux levels ten times the Crab flux, could be studied with a time resolution of seconds, under virtually background-free conditions \cite{bib:CTA}.

A fast reaction to unexpected transient \gray events is a crucial part of the CTA observatory, to trigger follow-up observations of astrophysical transients and better understand the origin of their emission. To capture these phenomena during their evolution and for effective communication to the astrophysical community, the speed is crucial and requires a system with a reliable automated trigger that can issue alerts immediately upon detection of $\gamma$-ray flares.  This will be accomplished by means of a RTA pipeline, a Science Alert System that will generate scientific alerts, becoming a key system of the CTA observatory.

The development and the implementation of the RTA system involves two Work Packages of the current CTA Product Breakdown Structure: the Data Management that has the responsibility of the RTA software development, and the array Control, that has the responsibility to provide the on-site Information and Communication Technology infrastructure (networking/computing architecture, cost estimations) to allow the RTA to successfully run.

\subsection{CTA requirements}

The CTA design imposes several key requirements to the RTA system: (i) scientific alerts must be generated with a latency of 30 s with respect to the triggering event collection; (ii) the search for transient phenomena must be performed on multiple timescales (i.e. using different integration time windows) from seconds to hours, both within a defined source region, and elsewhere in the field of view; (iii) the sensitivity of the analysis must be not worse than the one of the final analysis by more than a factor of 3; (iv) the availability of the RTA during observations must be greater than $98\%$.

\subsection{Technological constraints}

CTA is expected to generate a large data rate, due to the large number of telescopes, the amount of pixels in each telescope camera and the fact that several time samples (slices) are recorded in each pixel of the triggered telescopes. Depending on (i) the data reduction scheme adopted, (ii) the array layout, (iii)  the waveform acquisition (e.g. waveform sampling for selected pixels or for all pixels), (iv)  the trigger criteria, the estimations vary from 0,5 to 8 GB/s.
At time of writing the two sites where the arrays will be placed have not been chosen. In any case, it is reasonable to suppose that the candidate sites, especially for the CTA  South, will have a limited bandwidth for data transfer from the array site (North and South) to the Science Data Center (that receives the data from the arrays). For these reasons the CTA Observatory requires that the transfer of all acquired raw data from the two telescope sites is performed with a latency of less than 10 days and, in any case, the communications network to the sites must provide sufficient capacity to transfer event-level reconstruction parameters (but not the raw data) within 5 hours, for data from one night.

Due to the CTA requirement that the system should react with a latency of 30 seconds and due to the limited bandwidth between the telescope sites and the Science Data Center, the RTA should be performed on-site, i.e. with hardware and software infrastructures co-located with the telescopes; this means that this system is a part of the CTA On-Site Analysis.
\\
Thanks  to the large number of individual telescopes, CTA can be operated in a wide range of configurations, enabling operations with multiple sub-array targeting different objects or energy ranges, allowing the simultaneous monitoring of tens of potentially flaring objects. This will enhance the survey capability with respect to current instruments.

In addition, the CTA Observatory must be able to change the target (i.e. transition from data-taking on one target to data-taking on another target) anywhere in the observable sky within 90 s.

The coordination between the two sites and the real-time information transmission on the status and schedule of each observatory site are also desirable.

\section{The RTA as CTA key system}

In the CTA Observatory context the RTA is a key system for different reasons. It will be the only system that in real-time will enable CTA to:

\begin{itemize}
\item generate real-time scientific alerts to CTA Observatory itself and to other instruments in case of

\begin{itemize}
\item Gamma-Ray Bursts (GRB)
\item serendipitous discoveries in the Field of View of the array
\item detecting particularly interesting states about a given source under observation, with the consequent extension of the current observation if needed
\end{itemize}
\item provide real-time feedback to external received alerts.
\end{itemize}

The real-time science alerts generated by  the RTA system can change the short-term schedule of the Observatory. 

\section{A functional view of the RTA}

In this Section a short description of the functional view of the RTA is provided. To better understand the RTA functionality, Figure \ref{fig_logmodel} shows the functional schema of the overall On-Site Analysis with its context. 

\subsection{The context}

The \textit{Central Event Builder} merges the data from different telescopes triggered by the same event \cite{bib:hoff2012, bib:tej2013}. Its output (DL0 raw data, where DL means Data Level) is sent to the RTA and the \textit{Data Quality Analysis} (DQA). A temporary storage area (the \textit{DL0 Data Archive}) is foreseen. 

The \textit{Engineering Performance Analysis} (EPA) processes the data from the \textit{auxiliary items}, like camera housekeeping data, slow control data, weather station, etc., to identify emerging problems and ensure the reliable functioning of each CTA element. The EPA can generate engineering (ENG) alerts to RTA and DQA to avoid fake alert generations if some problems at hardware level occur. These alerts are also stored in a dedicated database for subsequent data retrieval (the \textit{ENG Alert DB}).

The \textit {Reconstruction} pipeline (RECO) is devoted to reconstructing the physical characteristics of the astrophysical \gray and background cosmic rays: it starts from the DL0 data generated by the Event Builder and produces the event lists which include the calibration, the gamma/hadron separation, and the energy and direction reconstruction of the incoming events using the calibration matrices. There are two types of RECO pipelines: the DEL-RECO (Delayed/Off-line-RECO) as part of the DQA, and the RT-RECO (Real-Time RECO) as part of the RTA. The primary products of both pipelines are the physical parameters of the \gray event along with the corresponding instrumental response data.  

The \textit{DQA} processes the data observation-by-observation and performs a scientific quality assessment of the scheduled observations starting from the intermediate and final results of the DEL-RECO  pipeline: this pipeline should be the same as off-site reconstruction pipeline but with a lower sensitivity due to on-site constraints, but if the on-site computer power will be sufficient the DEL-RECO should perform a full reconstruction of the events such as that performed off-site in the Data Center. The results of the DEL-RECO are stored into the \textit{DEL Data Archive}. The \textit{Data Quality Monitoring} provides the core functionalities of the DQA sub-system.

The \textit{Short-term scheduler}, that is part of the Central Array Control system, will promptly take all the actions needed to observe any RTA detected flaring source as well as any external Target of Opportunity request.

\subsection{The RTA}\label{rta_sec}

The RTA processes the data acquired by the Central Event Builder event-by-event during the telescope data acquisition. The \gray events are analyzed by this pipeline using standard or blind search methods to detect known or serendipitous astrophysical sources in high or unexpected state, select the best transient candidates and generate Science Alerts to provide a fast and automatic communication to the CTA community, to the  \textit{short-term Scheduler}, to the Flare Advocate and Array Operator. 

The RTA is divided into the following sub-functions:
\begin{enumerate}
\item the RT-RECO, the reconstruction pipeline that generates the event list, this should be a reduced version of the DEL-RECO with the main focus of the speed of analysis;
\item the Science Alert Monitoring (SAM), that detects unexpected astrophysical  events and generates Science Alerts;
\item the RT Health Monitoring, that performs a basic data quality check to evaluate the correct execution of the observations in real-time. This system generates alerts to SAM if the quality of the acquired data is not good; in this case the current observation is not analyzed;
\end{enumerate}

This pipeline receives  the ENG Alerts. RTA results should be stored in a database (the \textit{Science Alert DB}).

\begin{figure*}[t]
  \centering
  \includegraphics[width=0.9\textwidth]{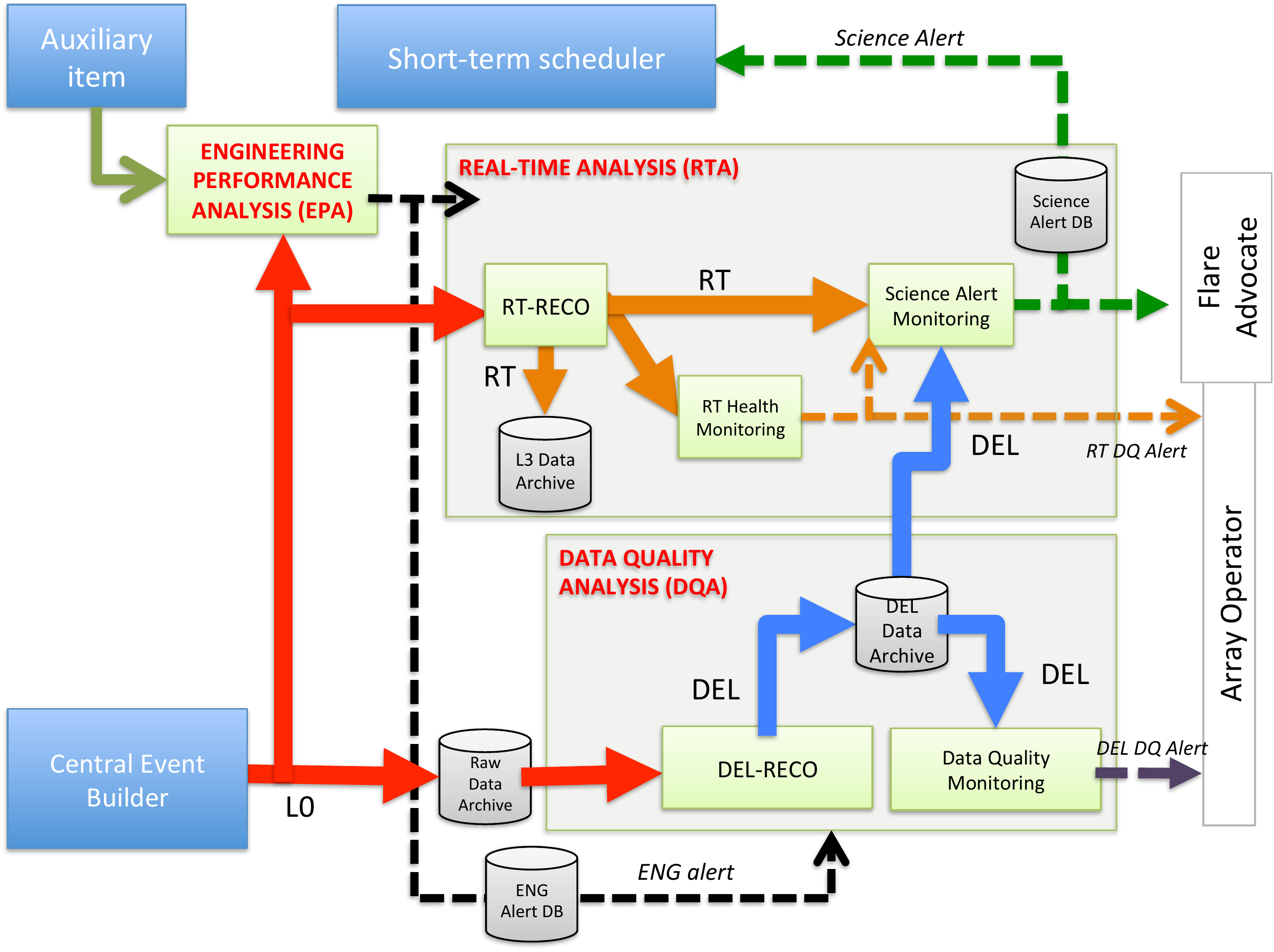}
  \caption{This functional logical view shows the CTA On-Site Analysis (the context of the Real-Time Analysis (RTA)), the basic functional blocks of the RTA, the data flow (continuous lines) and the alert flow (dashed lines). Engineering Performance Analysis (EPA) (under the array Control responsibility), Data Quality Analysis (DQA) and RTA (under the Data Management responsibility) are currently the main products of the CTA On-Site Analysis. }
  \label{fig_logmodel}
 \end{figure*}

In Figure \ref{fig_logmodel}, the link between the \textit{DEL Data Archive} and the \textit{Science Alert Monitoring} means that the last functional block should run  with a better reconstruction event list to reach higher sensitivity, but in this case the analysis is performed off-line (at the end of the observation).

 \section{A technological assessment}
 
 From a technological point of view the most stringent requirements of the RTA (the response time in conjunction with the required sensitivity and the fact that the system must operate in a wide range of sub-array configurations) impose a challenge in software and hardware solutions that should be adopted to fit these requirements, coupled with high data rate from telescopes and with the limited bandwidth between the telescope sites and the Science Data Center that suggests that  the analysis will be performed on-site. 
 
Performing the analysis on-site means that we will have limited electrical and computer power. However, substantial efforts in High Throughput Computing (HTC) are envisaged to enable efficient data processing.

For these reasons our working group has started the development of a technological assessment of the RTA. 

The main goal of this technological assessment is to test: (i) a set of frameworks and design patterns useful for the Inter Process Communication  between software processes running on memory; (ii) the sustainability of the foreseen CTA data rate in terms of data throughput, (iii) as a long term goal, how much we need to simplify  algorithms to be compliant with these requirements. 

The first step is the development of a simulator of the Central Event Builder, using the \textit{"Prod2"} simulated data \cite{bib:bern2013}. We convert these data into a stream of bytes \cite{bib:bulgarelli} that should be transmitted between the different processes of the RTA pipeline. 
The stream of byte  contains the data acquired from telescopes and will be formatted as a logical sequence of bits which represent well defined information, according with the definition of the satellite telemetry \cite{bib:pus}, basically to reuse a well defined standard.
Currently we are working on the sender and receiver processes.  

Given the very-demanding computational requirements for the RTA, modern accelerators - i.e., coprocessors - could come in handy to compute such a big data flow in real time. As a proof of concept, some tests carried out on Nvidia's Tesla GTX295 and Fermi C2075, where dummy content has been used, have shown data transfers through PCI-e 2.0 of 4.8 and 5.6 GB/s respectively in reading; 5.2 and 5.3
GB/s respectively in writing, using page-locked memory buffers. This data throughput between CPU and GPU memories is compatible with the foreseen CTA data rate. In addition, these figures should be doubled  if NVIDIA's Kepler cards,
which are PCI-e 3.0 compliant, are used.

In addition to that, several signal extraction algorithms \cite{bib:albert} are being currently ported to CUDA (Compute Unified Device Architecture), a parallel computing software development platform that provides extensions that enable expressing fine-grained and coarse-grained data and task parallelism, to assess their
performance since these cards are capable of concurrently running thousands of threads and possess a remarkable
throughput in floating-point computations.
For the short-term future, the porting of these algorithms to other API's (such as MPI, OpenCL, etc.) has been
planned.

\section{Conclusions}
The CTA requirements, the technological constraints and the foreseen observatory operations impose a challenge in the development of the RTA systems.  As a consequence, substantial efforts in the HTC are envisaged to enable efficient data processing, taking into account the limited electrical and computer power available on-site. A technological assessment of the RTA is in progress to test some solutions with current available technologies.

\vspace*{0.5cm}
\footnotesize{{\bf Acknowledgment:}{We gratefully acknowledge support from the agencies and organizations
listed in this page: http://www.cta-observatory.org/?q=node/22}}

\end{document}